\documentclass{article}
\usepackage[T1]{fontenc}
\usepackage[latin9]{inputenc}
\usepackage[letterpaper]{geometry}
\usepackage{textcomp}
\usepackage{amstext}
\usepackage{graphicx}
\usepackage{esint}
\usepackage[authoryear]{natbib}

\makeatletter
\newcommand{\lyxaddress}[1]{
	\par {\raggedright #1
	\vspace{1.4em}
	\noindent\par}
}

\makeatother

\begin{document}
\title{Saturn's deep atmospheric flows revealed by the Cassini Grand Finale
gravity measurements}
\author{Eli Galanti$^{1}$, Yohai Kaspi$^{1}$, Yamila Miguel$^{2}$, Tristan
Guillot$^{3}$, Daniele Durante$^{4}$, Paolo Racioppa$^{4}$,\\
 and Luciano Iess$^{4}$ \\
\\
(Published in GRL: http://dx.doi.org/10.1029/2018GL078087)}
\maketitle

\lyxaddress{\begin{center}
\textit{$^{1}$Department of Earth and Planetary Sciences, Weizmann
Institute of Science, Rehovot, Israel. }\\
\textit{$^{2}$Leiden Observatory, Leiden University, The Netherlands}\\
\textit{$^{3}$Observatoire de la Cote d\textquoteright Azur, Nice,
France}\\
\textit{$^{4}$Dipartimento di Ingegneria Meccanica e Aerospaziale,
Sapienza Università di Roma, Rome, Italy}
\par\end{center}}
\begin{abstract}
\textbf{How deep do Saturn's zonal winds penetrate below the cloud-level
has been a decades-long question, with important implications not
only for the atmospheric dynamics, but also for the interior density
structure, composition, magnetic field and core mass. The Cassini
Grand Finale gravity experiment enables answering this question for
the first time, with the premise that the planet's gravity harmonics
are affected not only by the rigid body density structure but also
by its flow field. Using a wide range of rigid body interior models
and an adjoint based thermal wind balance, we calculate the optimal
flow structure below the cloud-level and its depth. We find that with
a wind profile, largely consistent with the observed winds, when extended
to a depth of around $8,800$~km, all the gravity harmonics measured
by Cassini are explained. This solution is in agreement with considerations
of angular momentum conservation, and is consistent with magnetohydrodynamics
constraints.}
\end{abstract}

\section{Introduction{\normalsize{}\label{sec:Introduction}}}

\begin{figure}[t]
\centering{}\includegraphics[scale=0.8]{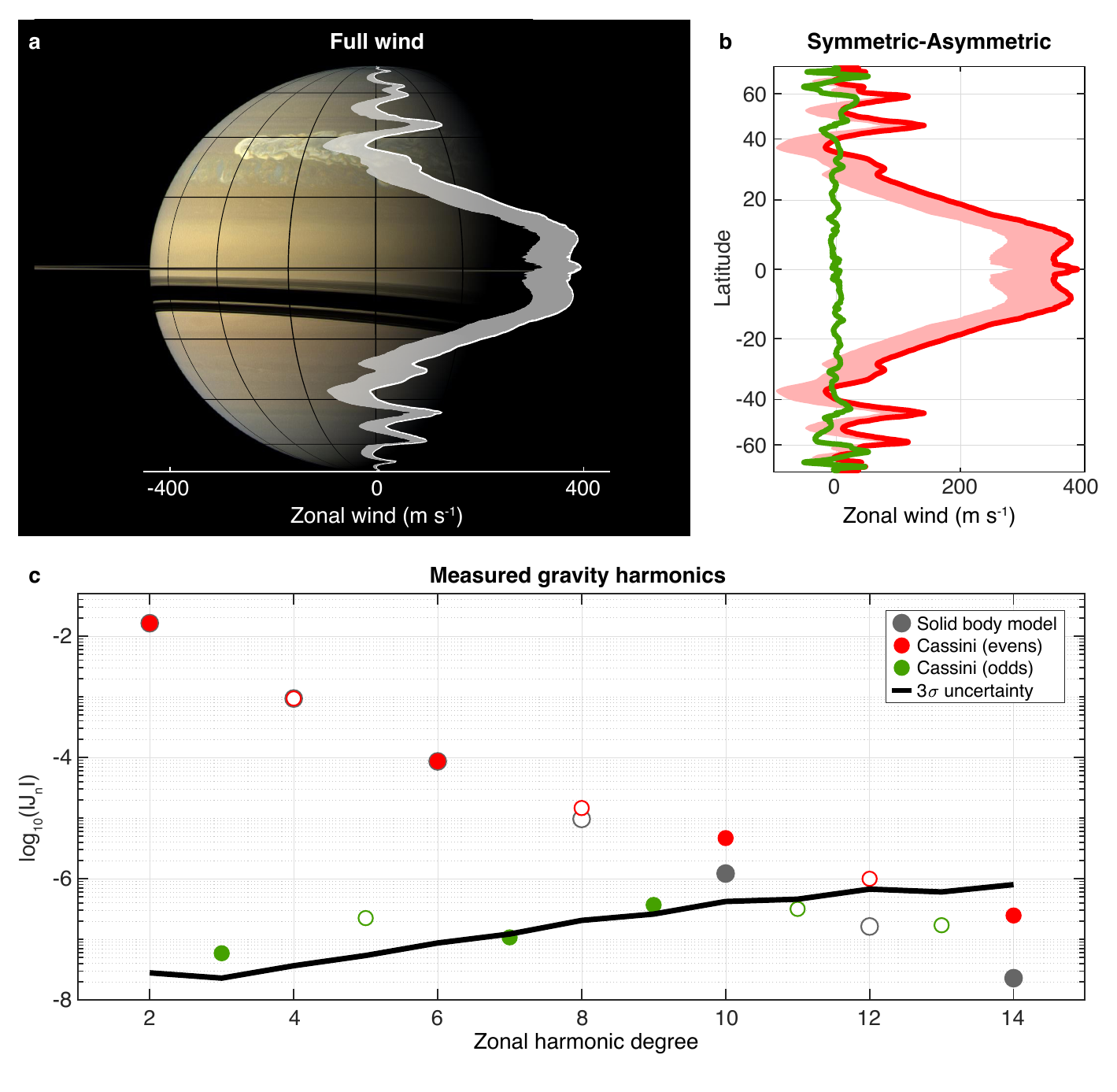}\caption{The Saturn's cloud-level winds (\citealt{Garcia-Melendo2011}, SB
channels). (a) The full zonal wind as function of latitude, placed
on top of a Cassini picture of Saturn (NASA/JPL-Caltech). Shown are
the winds based on the Voyager rotation rate (white), and the their
change if the \citet{Helled2015} rotation rate is used (gray area).
(b) The hemispherically symmetric (red) and asymmetric (green) components
of the wind. (c) The even and odd gravity harmonics measured by Cassini
(red and green dots, respectively), and the gravity field predicted
with a typical solid body model (gray dots). Also shown is the $3\sigma$
uncertainty associated with the measurements. \label{fig:wind-gravity-measurements}}
\end{figure}

Whether the fluid below Saturn's cloud levels is quiescent or exhibits
strong zonal flows has been a long lasting open question.
The zonal wind at the planet's cloud-level is well established based
on the Cassini measurements \citep{Garcia-Melendo2011}, with zonal
flows that reach $400$~m~s$^{-1}$ within a broad equatorial region,
and a few narrower jets at higher latitudes (Fig.~\ref{fig:wind-gravity-measurements}a,
white line). The flow is predominantly north-south hemispherically
symmetric (Fig.~\ref{fig:wind-gravity-measurements}b, red line),
with a much smaller asymmetric component that is more pronounced in
the mid to high latitudes (Fig.~\ref{fig:wind-gravity-measurements}b,
green line). 

However, aside from some observed variations in the wind strength
between the upper and middle troposphere, there has been very little
knowledge on the nature of the flow below the cloud-level. A possible
answer was enabled by the gravity experiment conducted during the
Grand Finale phase of NASA's Cassini spacecraft \citep{Iess2018}.
The measured gravity field (Fig.~\ref{fig:wind-gravity-measurements}c,
red and green dots) was found to be considerably different from that
predicted by typical internal rigid-body models (gray dots), especially
for gravity harmonics higher than $J_{6}$. Initial estimates for
the depth of the winds indicated very deep winds penetrating to a
depth of more than $9,000$~km \citep{Iess2018}. This estimate,
however, required a substantial modification of the meridional profile
of the cloud-level winds and was determined by matching only the gravity
harmonics $J_{3},\,J_{5},\,J_{8}$ and $J_{10}.$ In addition, the
background density profile used in the calculation of the wind-induced
gravity harmonics was based on a specific interior model that might
not represent all possible internal density structures.

was recently found that in Jupiter the part of the measured
gravity field unaccounted for with rigid body models, can be attributed
to a downward extension of Jupiter's cloud-level winds to a depth
of about $\sim3000$~km \citep{Kaspi2018,Guillot2018}.
This motivates consideration of whether a similar procedure works
for Saturn. While the flow inside a gas giant is expected to be aligned
parallel to the axis of rotation due to angular momentum constraints
(see \citealp[for a detailed discussion]{Kaspi2018}), it is not clear
whether this flow is well represented in the observed cloud-level
winds. The observations of Saturn's winds carry uncertainties that
need to be taken into consideration \citep{Garcia-Melendo2011}. First,
the sensitivity in the analyzed cloud-level winds reach $\pm20\,{\rm m\,}s^{-1}$
at certain latitudes. Second, it was found that there exists a substantial
wind shear between the wind at the upper ($\sim250$~mb) and middle
($\sim500$~mb) atmosphere, of up to $\pm100\,{\rm m\,}s^{-1}$ 
at the equator and up to $\pm20\,{\rm m\,}s^{-1}$
in the midlatitudes \citep{Garcia-Melendo2011}. A similar patten
was also found using thermal wind balance \citep{fletcher2008}, suggesting
that the variations represent changes with depth of the large scale
geostrophic flow. In addition, it was found that the flow exhibits
variations in time of up to $\pm50\,{\rm m\,}s^{-1}$
between the Voyager and Cassini observations. Therefore, it is possible
that the flow structure in the depths relevant to the gravitational
signal (thousands of kilometers deep) is somewhat different from that
observed at the cloud level.

Another uncertainty in the determination of the cloud-level winds
comes from of the need to calculate it with respect to the planet's
rotation rate, which is still not known with high certainty \citep{Helled2015}.
The estimate of \citet{Garcia-Melendo2011} (Fig.~\ref{fig:wind-gravity-measurements}a,
white line) was done with respect to the Voyager rotation rate \citep{Smith1982}.
More recent calculations \citep{Anderson2007,Read2009,Helled2015}
show a faster rotation rate that shifts the winds to more negative
values (Fig.~\ref{fig:wind-gravity-measurements}a, gray area). Note
that only the symmetric part of the wind is affected by the value
of the rotation rate (Fig.~\ref{fig:wind-gravity-measurements}b).
This uncertainty was found to have an effect on the shape and density
structure solutions of Saturn interior models \citep{Helled2013},
and a substantial effect on the wind-induced gravity harmonic $\Delta J_{2}$,
while having only minor effect on the higher wind induced even harmonics
\citep{Galanti2017d}.

In this study we aim to decipher the flow structure that best explains
all the gravity harmonics measured by Cassini in the gravity-dedicated
Grand Finale orbits. Unlike Jupiter, where large odd harmonics
were measured, in the case of Saturn we need to relay on the even
harmonics, requiring the calculation of the contribution of Saturn's
rigid body to the gravity field. We thus use a wide range of rigid
body (RB) models to map both the residual even gravity harmonics to
be explained by the flow, and to determine the preferable background
density profiles to be used in the calculation of the wind-induced
(WI) gravity harmonics. Using a thermal wind balance, we find the
top level zonal wind structure and its radial profile that best explain
the measured gravity field. This methodology provides a rigorous and
complete analysis of the gravity measurements.

The manuscript is organized as follows: in section~\ref{sec:RB-model}
we explore the RB solutions for the gravity field. These solutions
are then used to define the residual even gravity harmonics to be
explained by the flow, which is analyzed using the WI solutions (section~\ref{sec:TW-model}).
Next, we include in the WI gravity calculation a search for a top
level wind that allows the explanation of all measured gravity harmonics
(section~\ref{sec:modified-wind}). We discuss the results and conclude
in section~\ref{sec:Discussion-and-conclusions}.

\section{The rigid body (RB) gravity field\label{sec:RB-model}}

In order to explore the range of gravity solutions consistent with
the measurements we first construct interior models of Saturn that
fit the observational constraints of Saturn's radius and gravity harmonic
$J_{2}$, under the assumption that the interior of the planet rotates
as a rigid body (RB) following the methodology of \citet{Guillot2018},
which was effectively used for the analysis of Juno based Jupiter
gravity field. Since $J_{2}$ is also affected by differential
rotation \citep{Galanti2017d}, we allow the RB solution to deviate
from the measured value within the range $J_{2}^{{\rm RB}}=J_{2}^{{\rm obs}}\pm50\times10^{-6}$,
to cover a wide range of differential rotation scenarios corresponding
to different rotation rates and depths of the zonal flows. \citep{Galanti2017d}.

\begin{figure}[t]
\centering{}\includegraphics[scale=0.39]{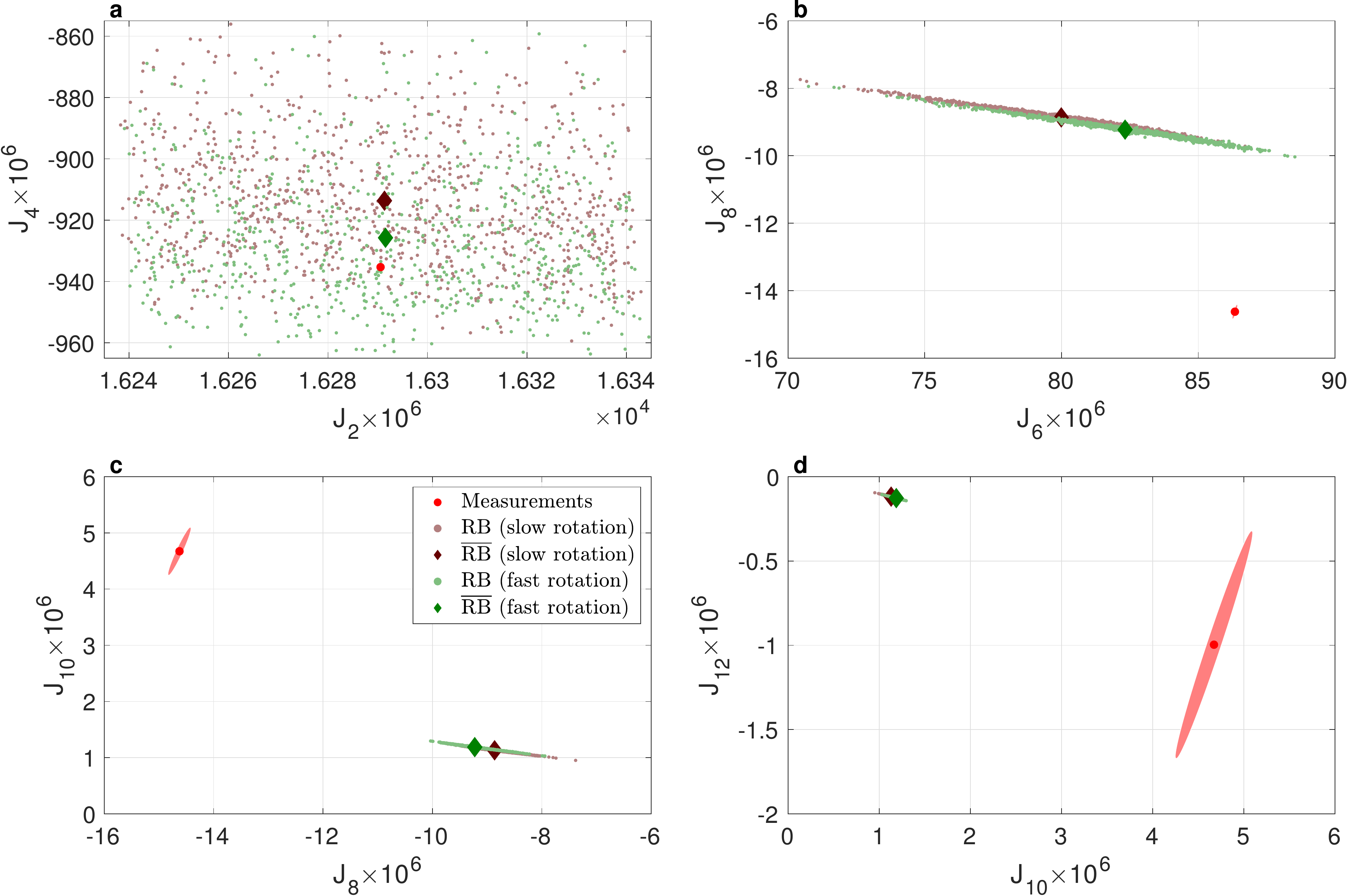}\caption{The even gravity harmonics ($\times10^{6})$ calculated with the RB
model based on slow rotation rate, $10{\rm h}~39{\rm m}~22{\rm s}$,
(blown dots) and fast rotation rate, $10{\rm h}~32m~45{\rm s}$, (green
dots), and their mean (brown and green diamonds, respectively). Also
shown are the Cassini measurements (red dots) and their $3\sigma$
uncertainties (red ovals).\label{fig:RB gravity field solutuions}}
\end{figure}

We assume a non-homogeneous structure for Saturn \citep[e.g., ][]{Guillot1999b,Fortney2003,Nettelmann2015,Vazan2016}
with a possible diluted core scenario, as proposed recently for Jupiter
\citep{Wahl2017}. The radial structure consists of 4 layers: (1)
an $H_{2}$ and $He$-poor atmosphere, (2) a metallic $H$ and $He$-rich
envelope, (3) a dilute core which is a metallic $H$ and $He$-rich
region with an increase in the heavy elements abundance, and (4) a
core composed by ices or rocks. The diluted core (layer 3) is omitted
in some of the models to allow exploration of both possibilities in
Saturn interior. We assume an adiabatic interior, neglecting the effects
of a non-adiabatic region on the gravitational harmonics, which are
significantly smaller than the uncertainties discussed in this paper
\citep{Nettelmann2015}. The outer boundary condition is set to $T=135\pm5\,{\rm K}$
at $1$~bar based on Voyager measurements \citep{Lindal1992}. The
abundance of Helium in Saturn's atmosphere is assumed to be $0.18\pm0.07$
\citep{Helled2013} and the value in the deeper layer is adjusted
so that the overall Helium abundance corresponds to the assumed protosolar
$He$ abundance of $0.270\pm0.005$ \citep{Bahcall1995}.
The $He$ phase transition is set to occur at a pressure between $1$
and $4$ Mbar, in agreement with immiscibility calculations \citep{Morales2013}.

A crucial parameter in the modeling of the internal structures of
giant planets is the equation of state \citep{Hubbard2016,Miguel2016}.
For $H$ and $He$ we use REOS3 \citep{Becker2014} as well as MH13
\citep{Militzer2013} that were derived using ab initio calculations,
with some adjustments as detailed in \citet{Miguel2016}. The heavy
elements, composed by rocks and ices, are modeled using the equations
of state for a mixture of silicates "dry sand"
for rocks and "water" for ices \citep{Lyon1992}.

In order to consider all possible interior structure configurations,
some of the parameters that are poorly constrained were chosen randomly
within a broad range (see supporting information). The simulations
were performed with either a slow $10{\rm h}~39{\rm m}~22{\rm s}$
rotation rate \citep{Smith1982} or a fast $10{\rm h}~32{\rm m}~45{\rm s}$
rotation rate \citep{Helled2015}. In total, a little over 1700 possible
interior models for Saturn were calculated.

The gravitational harmonics are calculated using the theory of figures
of 4th order \citep{Zharkov1978,Nettelmann2017}, combined with an
integration of the recombined density structure in two-dimensions
using a Gauss-Legendre quadrature \citep{Guillot2018}. This numerically
efficient method allows the execution of the hundreds of
calculations required. It has a known systematic bias compared
to more detailed calculations made with a Concentric Maclaurin Spheroids
method \citep{Hubbard2012,Hubbard2013}, therefore all gravity solutions
presented here are corrected with $\delta J_{2}=33\times10^{-6}$,
$\delta J_{4}=-3.816\times10^{-6}$, $\delta J_{6}=0.069\times10^{-6}$,
$\delta J_{8}=0.801\times10^{-6}$, $\delta J_{10}=-0.213\times10^{-6}$
and $\delta J_{12}=0.0045\times10^{-6}$, calculated similarly to
\citet{Guillot2018}.

The range of RB solutions for the even gravity harmonics are shown
in Fig.~\ref{fig:RB gravity field solutuions} for the slower rotation
rate (brown dots) and the faster rotation rate (green dots), together
with the measurements (red dots and ovals). Out of the 6 parameters
defined above, the He transition depth has the largest effect on the
solutions - larger transition pressure results in larger absolute
values of $J_{n}$ for $n>2$. Note that the parameters varied
in the $~1700$ RB models affect mostly the deep interior structure
of Saturn, hence $J_{2}$ and $J_{4}$, and have a weaker effect on
the atmosphere of Saturn, reflected in the higher even harmonics.
As a result, the solution dispersion is large for $J_{2}$ and $J_{4}$,
and get smaller for higher harmonics. Note also that the fast rotation
rate moves the solutions substantially toward the measurements, with
larger absolute values in all harmonics.But most importantly,
it is evident that while $J_{2}$ and $J_{4}$ can be explained entirely
by the RB models (the measurements are well within the model uncertainties),
the solutions for the higher harmonics, while closer to the measurements
and having a much larger spread than the model used in \citet{Iess2018},
cannot explain the measured values. The mean distance between the
model solutions (brown and green diamonds) and the measured values
(red dots) must to be the result of a wind-induced density anomalies.

\section{The wind-induced (WI) gravity field\label{sec:TW-model}}

\begin{figure}
\centering{}\includegraphics[scale=0.6]{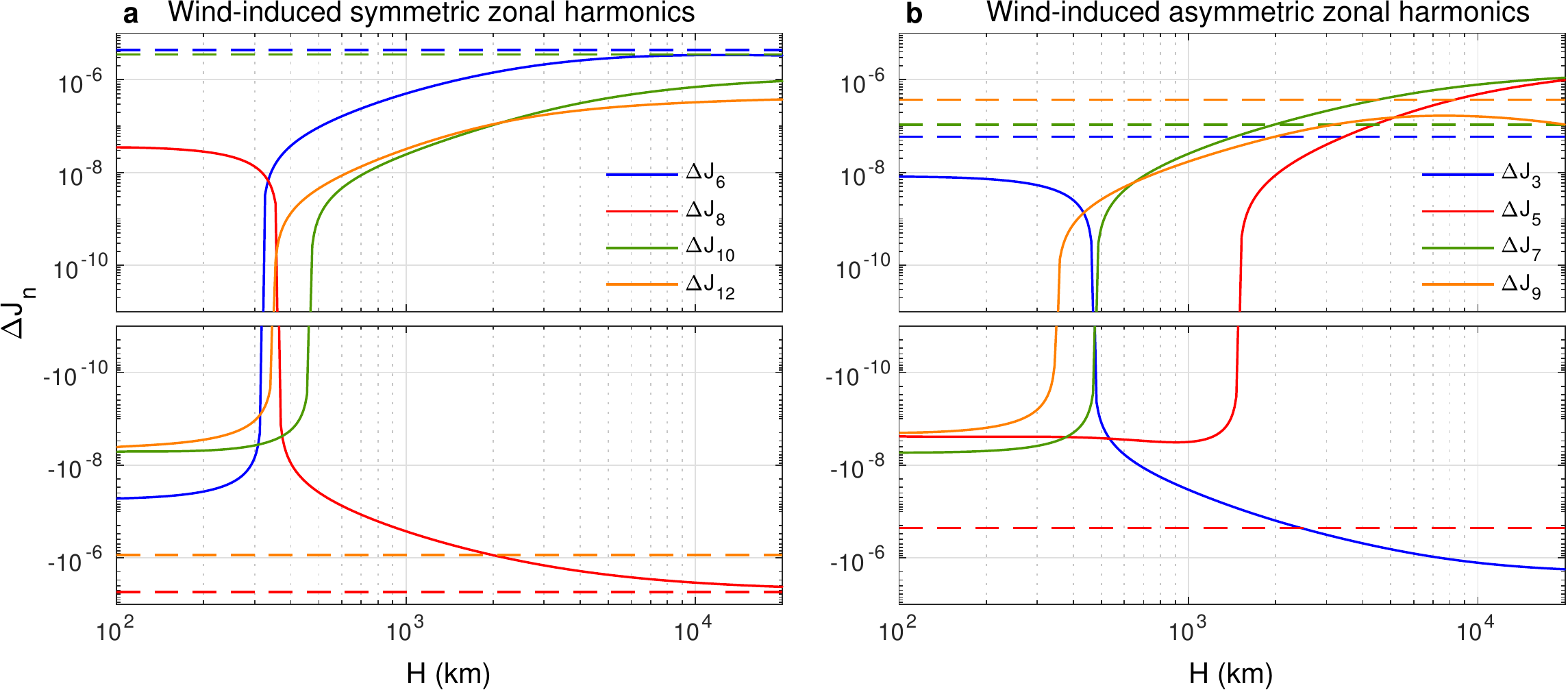}\caption{\label{fig:TW-prediction} The wind-induced gravity harmonics as predicted
by a thermal wind model with an exponential decay function, as function
of the e-folding depth $H$. (a) The even harmonics $J_{6},J_{8},J_{10}\,{\rm {\rm and}}\,J_{12}$.
(b) The odd harmonics $J_{3},J_{5},J_{7}\,{\rm {\rm and}}\,J_{9}$.
Also shown are the measured harmonics (dashed lines) of the evens
calculated from Fig.~\ref{fig:RB gravity field solutuions}, and
the odds taken directly from the measurements.}
\end{figure}

Given that Saturn is a large planet and a fast rotator, any large
scale flow is governed by the thermal wind balance, relating the flow
to the density field \citep{Pedlosky1987,Kaspi2009}. Assuming the
flow is zonally symmetric and assuming sphericity \citep{Galanti2017a},
the dynamical balance is between the flow gradient in the direction
parallel to the axis of rotation and the meridional gradient of density
perturbations
\begin{equation}
2\Omega r\frac{\partial}{\partial z}\left(\rho_{0}u\right)=g_{0}\frac{\partial\rho'}{\partial\theta},\label{eq: thermal wind}
\end{equation}
where $\Omega$ is the planet's rotation rate, $\rho_{0}(r)$ and
$g_{o}(r)$ are the rigid body density and gravity fields, $u(r,\theta)$
is the flow field, $\rho'(r,\theta)$ is the anomalous density field,
and $r,\theta$ and $z$ are the radial, latitudinal, and axis of
rotation directions (see supporting information for a detailed
derivation). This balance was used extensively to study the wind
structure on Jupiter \citep[e.g.,][]{Kaspi2013a,Liu2013,Zhang2015,Kaspi2016,Galanti2017a},
as well as for the prediction of the wind-induced gravity field to
be expected on Saturn \citep{Kaspi2013a,Galanti2017d}. The
wind-induced (WI) gravity harmonics are calculated as the volume
integral of $\rho'$ projected onto Legendre polynomials
\begin{equation}
\Delta J_{n}^{{\rm }}=\frac{2\pi}{MR_{e}^{n}}\intop_{0}^{R_{e}}r{}^{n+2}dr\intop_{\theta=-\pi/2}^{\pi/2}P_{n}\left(\sin\theta\right)\rho'\left(r,\theta\right)\cos\theta d\theta,\label{eq: Jn model}
\end{equation}
where $M$ is the planetary mass, $R_{e}$ is the planet equatorial
radius, and $P_{n}$ are the Legendre polynomials. For a detailed
discussion of the method refer to \citet{Kaspi2018}.

\begin{figure}[t]
\centering{}\includegraphics[scale=0.48]{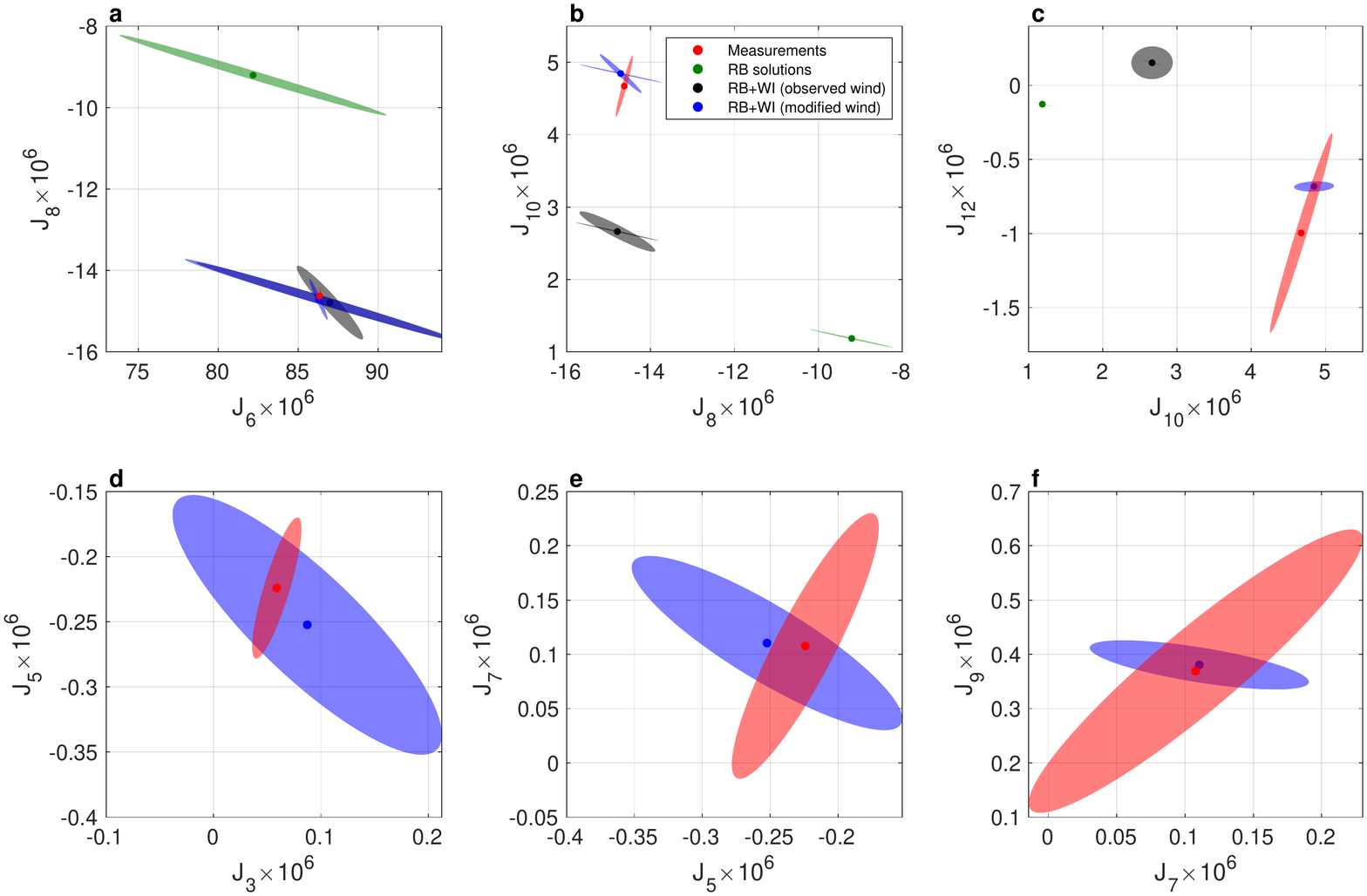}\caption{\label{fig:gravity-solutions} The the gravity harmonics ($\times10^{6})$
calculated using the RB model only (green), RB combined with the WI
based on observed winds (gray), and RB combined with WI based on optimal
winds (blue). Also shown are the measured values (red). Upper panels
show the even harmonics $J_{6},J_{8},J_{10}\,{\rm {\rm and}}\,J_{12}$.
Lower panels show the odd harmonics $J_{3},J_{5},J_{7}\,{\rm {\rm and}}\,J_{9}$,
which have no RB solutions. The uncertainties of the RB solutions
are calculated from the range presented in Fig.~\ref{fig:RB gravity field solutuions}.
For the WI solutions we show both the uncertainties resulting from
the RB solutions and those associated with the WI solutions.}
\end{figure}

Following the same methodology used in \citet{Kaspi2018} we first
assume the wind decays exponentially with a decay depth $H$ (see
supporting information), and calculate the resulting gravity harmonics
$\Delta J_{n}$ as function of $H$ (Fig.~\ref{fig:TW-prediction},
solid lines). Given the mean values of the even harmonics of the RB
model (Fig.~\ref{fig:RB gravity field solutuions}, green diamonds),
we can now plot the measured $\Delta J_{n}$ (Fig.~\ref{fig:TW-prediction}a,
dashed lines) in addition to the measured odd harmonics Fig.~\ref{fig:TW-prediction}b,
dashed lines), which are not affected by the RB solutions. Based on
the observed cloud-level winds, this model is able to partially explain
the even harmonics $J_{6},J_{8}$ and $J_{10}$, for which
it asymptotically reaches the measured values of $J_{6}$ and $J_{8}$,
and about one third the value of $J_{10}$, but cannot explain the
odd harmonics. The WI solutions are found to be effected by the background
density $\rho_{0}$ taken from the RB solutions - profiles with the
highest densities in the outer layers ($r>0.7R_{s}$) increase the
value of the gravity harmonics by up to $50\%$, compared
to RB profiles with the lowest values.

We can then expend our search with a more complex decay function (see
supporting information), looking for the optimal radial structure
of the wind that gives the best fit the measured gravity harmonics.
The optimal solution is found with $H_{0}=11,547\pm875\,{\rm km}$,
$\Delta H=1100\pm308\,{\rm km}$, and $\alpha=0.89\pm0.11$,
where $H_{0}$ is the depth of a hyperbolic tangent and exponential
radial profiles, $\Delta H$ is its width and $\alpha$ is the ratio
between the two functions. The resulting gravity harmonics for $J_{6},J_{8}\,{\rm and}\,J_{10}$
are shown in Fig.~\ref{fig:gravity-solutions}a,b,c as (black dots)
along with the uncertainties associated with them (gray ovals). Also
shown are the uncertainties associated with the RB solutions (additional
gray ovals). As predicted with the simple model (Fig.~\ref{fig:TW-prediction}),
the optimal $J_{6},J_{8}$ and $J_{10}$ are able to move the RB solution
(green dots and ovals) in the direction of the measurements (red dots
and ovals), with $J_{6}$ and $J_{8}$ explained within the uncertainties
associated with the model, and $J_{10}$ being pushed half
the way to the measurements. The odd harmonics remain inconsistent
with the measurements and are outside the range presented in the figure.
This implies that the observed cloud-level wind profile might not
represent accurately the flow affecting the gravity field.

\section{The wind below the cloud-level\label{sec:modified-wind}}

The limited ability to explain all the measured $\Delta J_{n}$ when
using the observed cloud-level wind suggests that the wind-induced
gravity signal might be a result of a flow field that is somewhat
different from that observed at the cloud-level. As discussed in section~\ref{sec:Introduction}
this possibility has support in the observations, in which uncertainties
in the analysis as well as variations in both time and depth are observed
\citep[Figs. 3 and 9]{Garcia-Melendo2011}. This implies that the
search for a flow field that explains the measured gravity field requires
an augmented optimization, one that would also allow the cloud-level
wind itself to vary in addition to its radial decay profile. This
can be achieved by decomposing the cloud-level wind into the first
$N$ Legendre polynomials

\begin{eqnarray}
U_{{\rm }}^{{\rm sol}}(\theta) & = & \sum_{i=0}^{N}A_{i}^{{\rm sol}}P_{i}(\sin\theta),\label{eq:U-reconstruction}
\end{eqnarray}
where $A_{i}^{{\rm sol}}$ are the coefficients defining the meridional
wind solution. The optimization procedure for calculating
$A_{i}^{{\rm sol}}$ is constructed to ensure that the deviation of
the wind solution from the observed cloud-level wind is not larger
then what is necessary to bring the gravity field solution within
the uncertainty range of the measured field (see supporting information).

\begin{figure}
\centering{}\includegraphics[scale=0.43]{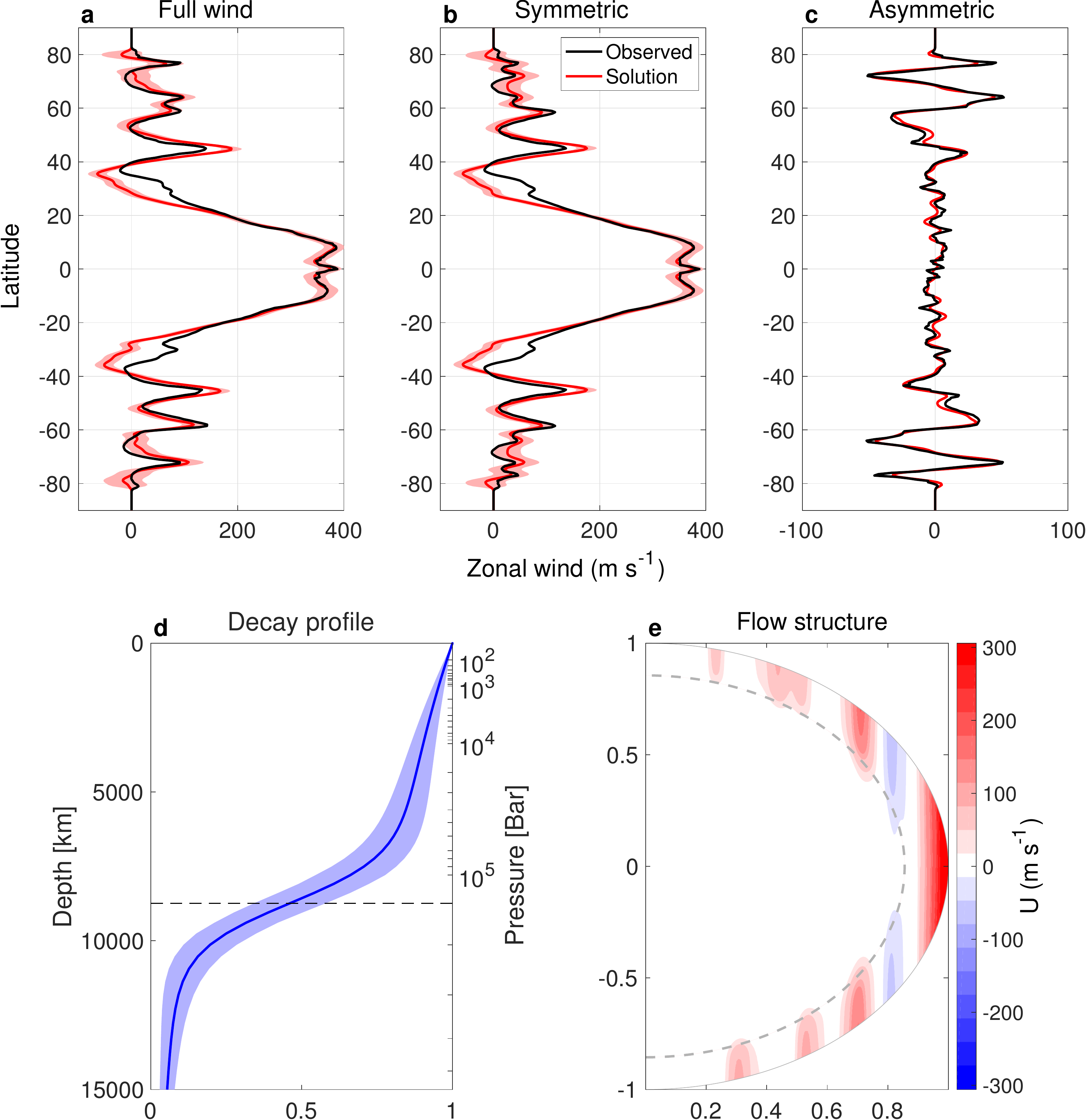}\caption{\label{fig:wind-solutions} The model solution with modified
cloud-level wind. (a) the observed cloud-level winds (black) and
the optimized wind profile (red lines), together with the uncertainty
(red shading) associated with the model solution (blue shadings in
Fig.~\ref{fig:gravity-solutions}). (b) the symmetric part of the
winds, and (c) the asymmetric part of the winds. (d) the optimized
radial decay of the winds as function of depth and pressure, with
its uncertainty in blue shading. (e) The flow structure resulting
from the model solution for the cloud-level wind (red line in panel
a), projected inward in the direction of the spin axis and decaying
radially according to the profile in panel d. The dashed line in both
(d) and (e) denotes the depth of 8,743~km.}
\end{figure}

The optimal solution for the radial structure of the flow is found
with $H_{0}=8,743\pm102\,{\rm km}$, $\Delta H=1577\pm1435\,{\rm km}$,
and $\alpha=070\pm0.129$. The wind solution is shown in
Fig.~\ref{fig:wind-solutions}a-c (red lines), together with its
uncertainty (red shading). Also shown is the observed cloud-level
wind (black lines). In most latitudes the solution wind is very similar
to the observed cloud-level wind and is well within the expected uncertainties
discussed in Sec.~\ref{sec:Introduction}). The largest deviations
are around latitudes $25^{\circ}-35{}^{\circ}$ north and south, similar
in location to the findings of \citet{Iess2018}, but about half the
size. The wind decay profile and the resulting flow structure
(Fig.~\ref{fig:wind-solutions}d-e) reveal that the wind behaves
nearly barotropicly in the equatorial region (extending all the way
to the equatorial plain in the direction of the spin axis), but nearly
baroclinicly outside latitudes $20^{\circ}N$ and $20^{\circ}S$,
i.e. decaying before reaching the equatorial plain.

With the modified wind the WI model is able to fit all gravity harmonics
taken into consideration (Fig.~\ref{fig:gravity-solutions}, blue
dots and ovals), both the even and the odd harmonics. Importantly,
the goal here is to have an overlap between the uncertainty of all
the model gravity harmonics solutions (blue ovals) and the measurement
uncertainties (red ovals). It would have been easy to get the model
solutions (blue dots) to fit exactly the measurements (red dots),
simply by relaxing the regularization of the winds $\epsilon_{U}$
(see supporting information), but with the cost of the wind
solution getting farther away from the observations. By aiming for
an overlap of the uncertainty ovals only, a balance is reached between
the need for a viable solution and the need to keep the optimized
winds as close as possible to the cloud-level observations.

\section{Discussion and conclusion\label{sec:Discussion-and-conclusions}}

The Cassini gravity measurements provide a unique opportunity to decipher
the nature of the flow on Saturn. The initial analysis of \citet{Iess2018}
pointed to the existence of deep flows in the equatorial region, yet
these results were limited to a specific rigid body model, and as
a result, required the surface wind to be substantially different
from the observed cloud-level winds.

In this study we investigate the gravity field of Saturn, using a
wide range of rigid body (RB) gravity models and a wind-induced gravity
model in which the top-level wind is allowed to differ from the observed
cloud-level wind. The RB solutions, while having a much broader range
of solutions than those used in \citet{Iess2018}, are still distinctively
different from the measurement for all even harmonics higher than
$J_{4}$, therefore implying the existence of strong differential
flows underneath the cloud-level. They also exhibit a considerable
variance in the radial density structure, with higher densities in
the outer layers associated with up to $50\%$ higher values of $J_{6},\,J_{8}$
and $J_{10}$, compared with the values obtained with the
lower background density. These cases, explaining better the measurements,
are mostly associated with a deeper $He$ transition depth ($P\approx4\,{\rm Mbar}$).
Interestingly, these cases also allow the wind-induced gravity signal
to match the measurements with less modification of the cloud-level
wind, since both the RB and the WI solutions have higher values of
$J_{6},J_{8},J_{10}$ and $J_{12}$, thus pushing their sum farther
toward the measurements.

With a conservatively modified cloud-level wind, extended to a depth
of around $8800$~km, all the relevant gravity harmonics can be explained,
taking into account the associated uncertainties in both the measurements
and the model solutions. In most latitudes the optimal top level wind
is similar to the observed wind, with the largest deviations found
around latitudes $25^{\circ}-35{}^{\circ}$ north and south, similar
to the deviation found by \citet{Iess2018} but twice as small. This
is a result of the different SB background density and gravity harmonics
solutions used here, as discussed above. In order to explain the
measured odd harmonics, the modifications needed in the asymmetric
flow are minor and are well within the uncertainty in the cloud-level
wind observations.

The optimal flow depth of $8800$~km is consistent with estimates
of the depth in which the conductivity of the fluid prohibit strong
flows \citep{Liu2008,Cao2017}. Interestingly, this depth, taken at
the equator, is approximately the location where the cylindrical flow
outcropping at $25^{\circ}-35{}^{\circ}$ is, and where the flow is
found to be most different from that observed at the cloud-level.
Investigating this circumstantial result might require the analysis
of a dynamical model in which magnetohydrodynamical considerations
are included \citep{Galanti2017e}.

\textit{Acknowledgments:} EG and YK  acknowledge support from the Israeli Space Agency and from the Helen Kimmel Center for Planetary Science at the Weizmann Institute of Science. DD, PR, and LI were supported by the Italian Space Agency. The data plotted in the figures is available at https://www.dropbox.com/sh/enef9c2z1lb5x1s/AAByea2T9Cx-oqtLYOfuvCjsa?dl=0.

%\bibliographystyle{agu}
%\bibliography{../../../Manuscripts/bibliography_main}

\end{document}